\documentclass[conference]{IEEEtran}
\IEEEoverridecommandlockouts
\usepackage{cite}

\usepackage[dvipsnames]{xcolor}
\usepackage{amsmath,amssymb,amsfonts}
\usepackage{algorithmic}
\usepackage{graphicx}
\usepackage{textcomp}
\usepackage{xcolor}
\usepackage[]{algorithm2e}
\usepackage{caption}
\def\BibTeX{{\rm B\kern-.05em{\sc i\kern-.025em b}\kern-.08em
    T\kern-.1667em\lower.7ex\hbox{E}\kern-.125emX}}
\begin{document}

\title{SARiSsa - A Mobile Application for the Proactive Control of SARS-CoV-2 Spread}

%
%
%
%
%
%
%
%
%
\DeclareRobustCommand*{\IEEEauthorrefmark}[1]{%
  \raisebox{0pt}[0pt][0pt]{\textsuperscript{\footnotesize #1}}%
}

\author{\IEEEauthorblockN{Christos~Chondros\IEEEauthorrefmark{},
Christos~Georgiou-Mousses\IEEEauthorrefmark{}, Stavros~D.~Nikolopoulos\IEEEauthorrefmark{},\\ \bigskip
Iosif~Polenakis\IEEEauthorrefmark{} and Vasileios~Vouronikos\IEEEauthorrefmark{}}
\IEEEauthorblockA{\it Department of Computer Science and Engineering \\
 University of Ioannina\\
Ioannina, Greece\\}}

\maketitle

\begin{abstract}
In this work we propose the design principles behind the development of a smart application utilized by mobile devices in order to control the spread of SARS-CoV-2 coronavirus disease that caused the COVID-19 pandemic. Through the deployment of this application utilizing their Bluetooth enabled devices, individuals may keep track of their close contacts, and if nearby contacts using the same application are reported later as infected the proximate individual is informed in order to be quarantined for a short of time, preventing hence the spread of the virus. Through the latest year, there have been developed several applications in the Google Play Store that can be deployed by smart devices utilizing their Bluetooth connectivity for the nearby device tracking. However, in this work we propose an open architecture for the development of such applications, that also incorporates a more elaborated graph-theoretic and algorithmic background regarding the contact tracing. The proposed contact tracing algorithm, that can be embedded in the deployment of the application, provides a more immediate tracking of the contacts of an infected individuals, providing a wider extent in the tracing of the contacts, leading hence to a more immediate mitigation of the epidemic.
\end{abstract}


\section{Introduction}
In this section we discuss the effect of the epidemic mitigation and how the deployment of new technologies contributes to the pandemic prevention, reducing hence the potential life-costs may caused by a spreading disease, present some of the most used mobile applications developed for the deployment of the procedure of contact tracing regarding the COVID-19 epidemic, and propose the basis beyond the construction of an open architecture for the deployment of such applications alongside the contribution provided by the theoretical background upon which the proposed architecture is developed on.

	\subsection{Epidemic Control}
	The study of the epidemics has been of major importance, especially the latest year that the new coronavirus SARS-CoV-2 that caused the COVID-19 pandemic  appeared costing thousands of lives leading to a further extent to several social and economical issues. Regarding airborne pathogens, they can be transmitted when two people are at a certain distance and there is an interaction between them, e.g., speech, coughing, sneezing, which spreads a quantity of infected droplets. Epidemic outbreaks can be occur through the population, or they can remain low for long periods of time, experiencing sudden outbreaks or even patterns of wave increase and decrease in host proliferation. The main features of an epidemic are the frequency of its occurrence, its duration and its evolution among individuals in a population. This information can be further used to improve health services as well as to study the prevention and control of the progression of possible outbreaks of pandemics.
	
	\subsection{Related Work}
One of the most important aspects in the reduction of the spread of SARS-CoV-2 disease is the control of the transmission mostly among communities that expose a high risk \cite{GuBeGaLoBiAn}, as also in the wider geographically generated communities that may occur through  the extent of a geographical area in terms of congestion. For the later case, several approaches that leverage the utilization of short range communication technologies, such as Bluetooth, have been proposed leading to the development of smart mobile applications that focus mainly on the tracing of contacts, mostly of infected individuals, made during the last days. The latest mobile applications proposed and deployed, mostly for Android devices, as the {\tt Corona Warn App}~\cite{coronawarnapp}, the {\tt Covid Out}~\cite{covidout}, the {\tt Covid Safe}~\cite{covidsafe}, the {\tt Covid Tracer}~\cite{covidtracer}, the {\tt Covid Tracker}~\cite{covidtrackerIR}, the {\tt Coronalert}~\cite{coronalert}, and the {\tt StayAway} ~\cite{stayaway}, where the vast majority of them utilize the Bluetooth technology and interaction handshake among the nearby Bluetooth enabled devices in order to exchange the required information regarding the identity of each device, create a contact list of the devices a user has been daily in contact augmenting thus the contact tracing procedure by registering them and informing the user whether one or more of them have been declared as infected.

Bluetooth based smart applications that focus on contact tracing have received an extended review over the aspect of acceptance from the public, in terms of how much suuch technologies are adopted by the users \cite{CoGr,WyHo,OsRa}, where additionally, the main concern throughout every approach as it is described throughout the literature, is the security of the stored data \cite{BeJaYuIpJaPiSh, FiMaPiRaZh, HoHiMoVa,  ShLoShRa, SuWaXuTyCaRa, CoGr}, and how potential conflicts to the regulation may arise \cite{JaLa}. Data privacy and confidentiality should definitely be of main concern for such applications that utilize the sensitive and personal health data of individuals, while some approaches \cite{ViThBaDi,BeepTrace} have been proposed that deal with such issues.

\subsection{Contribution}
Through this work we propose the development of a mobile application, namely SARiSsa \cite{sarissa}, to reduce the spread of SARS-CoV-2. The basic principles of this system's integration are described and the corresponding requirements are fulfilled providing an adequately functional  mobile applications development aiming to reduce the extent of COVID-19 pandemic. The proposed approach provides a proactive prevention aspect where in contrast to the approaches proposed so far, it is designed to prevent individuals from being infected when being either in crowded places or even in cases that there exists a high risk of infection even in presence of a few infected individuals, rather than informing about an infected contacts they had they previous days or preforming only the traditional contact tracing procedure. This approach could lead to the development an even more effective strategy against the spread of SAR-CoV-2, satisfying also the demands concerning the confidentiality of sensitive data and the anonymity of the individuals that use the proposed application. Finally, regarding the augment of the contact tracing procedure we propose an efficient contact tracing algorithm that provides a more immediate tracking of the contacts in the present day where an individual has been detected as infected adding more individuals to quarantine but reducing on the other hand more drastically the spread of the new coronavirus or if applicable to any other airborne spread disease. To this end, it is worth noting that the proposed architecture, as also the corresponding application alongside the incorporated algorithmic background, are characterized by their abstraction and the simplicity on their deployment, retaining the ability to be adapted on later potential epidemics caused by airborne pathogens, or even pathogens which their transmission requires close contact of individuals.

\section{Proposed Model}
\label{sec:headings}
In this section we present the proposed model for the construction of an open architecture proposed and utilized for the development of SARiSsa mobile application deployed for contact tracing, regarding the underlying components and the overall system interaction, taking into account the properties of the spread of SARS-CoV-2 coronavirus, alongside the aspects that depict the utilization of the developed application with respect to its usability in real life scenarios.

	\subsection{System Architecture}
The overall architecture of our system is based upon the interaction of a standalone application that is located in the mobile device of each individual, and another application that is deployed by the verified health centers in order to register the users in the application, where all the data are hosted inside a data-base with all the sensitive information being either hashed or encrypted, ensuring the maximum standards on their confidentiality.

\begin{figure*}[t!]
\centering
		\includegraphics[scale=0.88]{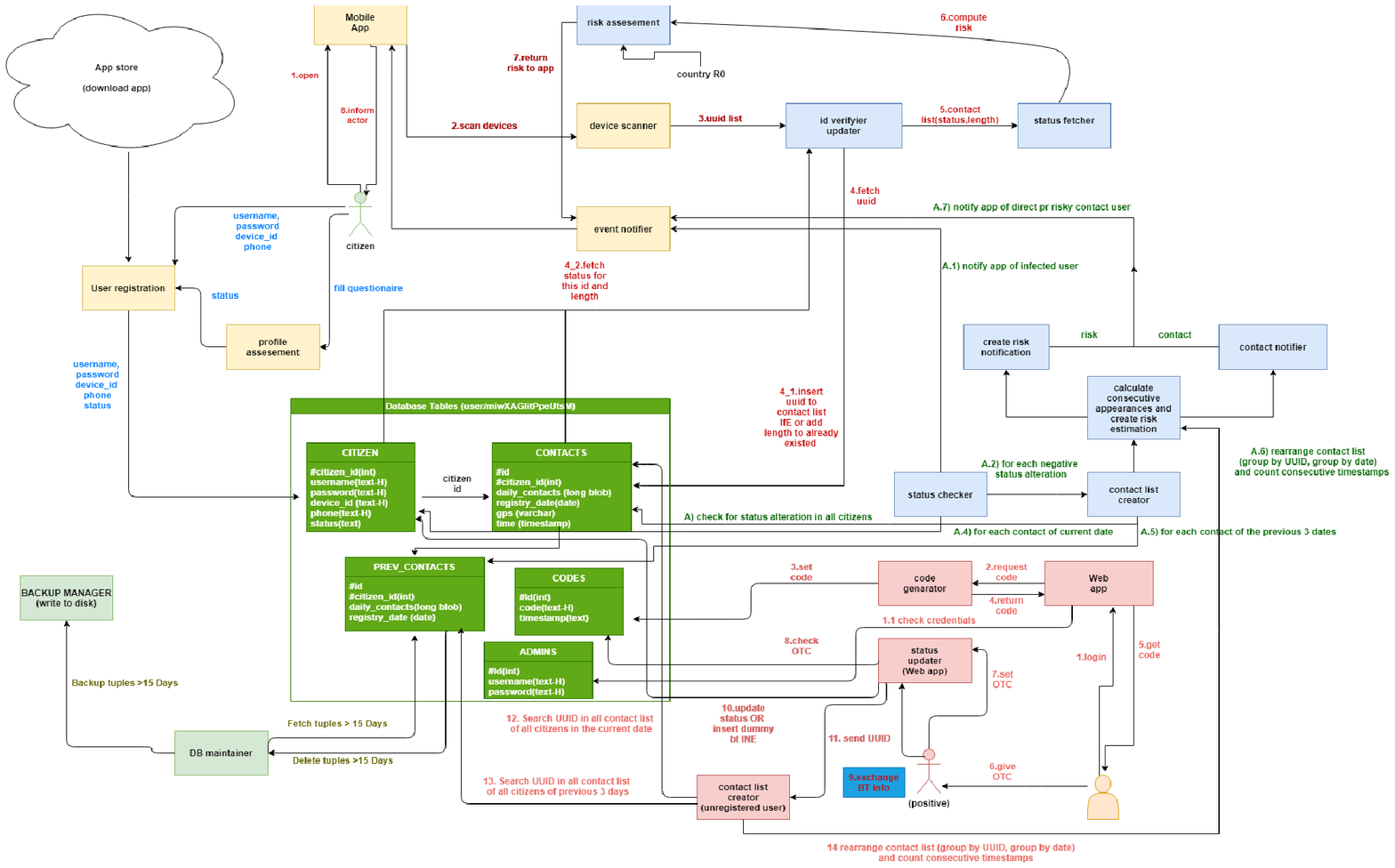}
	\caption{Integrated Architecture Scheme of SARiSsa Mobile Application Deployment.}
	\label{fig:fig1}
\end{figure*}

\bigskip
\noindent \textbf{Web-based Application.} The web-based application that is deployed by the authorized medical staff in health centers achieves a dual purpose, firstly on registering securely the new users to the system's data-base, as also, and most importantly, to register the status update of a user if the corresponding individual has been detected as ``positive" (i.e., infected by the SARS-CoV-2 coronavirus) after a probable antigen test. Throughout this procedure, the authorized health staff is logged into the system with specific credentials and asks for a ``One-Time-Code" (or, for short, OTC) in order to provide it to the user for his registration. The code generator module generates and returns this code, while it also inserts this code to the data-base in order to not be utilized again in the future. 

On the response of this procedure, the patient that is either infected, or not, enables the Bluetooth device, and when traceable, the corresponding Bluetooth info is transmitted to the health staff's device and then is registered to the system using the OTC ensuring the freshens of its credentials ensuring also that the user has not been already registered from a previous process, while in the case where the individual is infected, notifying recursively also the devices that include this Bluetooth info (i.e, the {\tt UUID} of the infected individual). 

\bigskip
\noindent \textbf{Mobile Application.} The mobile application, that is deployed by the individuals, has as its main objective to locate nearby Bluetooth enabled devices and record them into a list of contacts (i.e., {\it contact list}) with respect to the time a device has been for a specific time period at a specific distance from the individual's device, that in terms of disease transmission is interpreted as an increased probability of infection in case the owner of the nearby device is infected. Moreover, the client side of the mobile application is based mostly into two procedures, namely, the Device Scanner that detects the nearby Bluetooth enabled devices exchanging the corresponding {\tt UUID}s that uniquely identify each Bluetooth enabled device, and the Event Notifier, that informs the user of the application if one of its previous contacts has recently altered its status to infected, i.e., an update has been performed into the data-base by an authorized member of the medical staff through the corresponding Web-based application. 

Concerning the server sight internals that operate through the procedure described above, another module, the ID Verifyier Updater inserts the corresponding Bluetooth information gathered from the nearby devices into the list of contacts for each individual and indirectly, evaluating the status of these devices (i.e., if one or more  {\tt UUID}'s correspond to an infected or a contact of an infected user) computes a risk value through the Status Fetcher and Risk Assessment modules, informing the user of the application about the safety of remaining at the same location according to the surrounding contact's status.

\bigskip
\noindent \textbf{Proactive Protection.} Through this aspect, it is quite notable to refer that a proactive approach on the information of individuals is deployed in order to avoid proactively the infection rather than the usual approach deployed from the applications proposed so far that inform the user only after one of its contacts has been declared as infected. Through this approach it is achieved a further reduction to the numbers of infected individuals rather than reducing only the spread of the disease by tracing the contacts of an infected individual. 

Moreover, as we mentioned before, the user's data stored through the application are all either hashed or encrypted, however, it is worth noting that regarding the confidentiality of data, the user of the application has not the ability of detecting whether a specific device belongs to an infected user, accessing hence sensitive health data, but instead, the user is notified only about the risk of staying in this area with respect to number of nearby devices that belong to infected individuals. 

\begin{figure*}[t!]
\centering
		\includegraphics[scale=0.8]{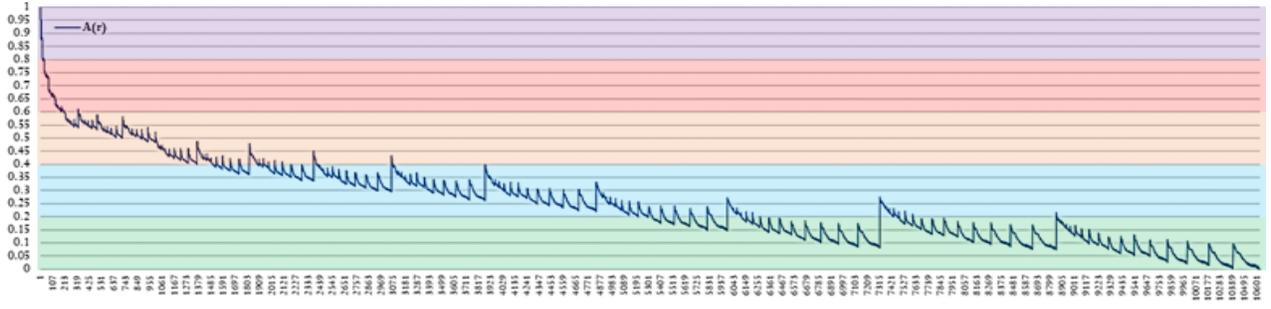}
	\caption{Study on the behavior of the risk assessment $A(r)$ for different combination of individual categories.}
	\label{fig:fig2}
\end{figure*}

Additionally, in a similar characterization of the risk exposed by the congestion of each area regarding the number of nearby Bluetooth enabled devices detected to be owned by registered individuals being in infected status, or even registered as contacts of infected individuals, our approach through the development of SARiSsa application is to characterize also the risk of the surrounding area exhibited by the status of nearby individuals, note that the status of an individual regarding  whether being characterized as infected, can only be set by the authorized medical staff. However, potentially risky individuals, i.e., individuals that are close contacts of individuals that have been detected as infected, or individuals that are contacts (or close contacts) of contacts of individuals that are infected, or even individuals that usually are gathered and remain for long time periods in  crowded and congested places that exhibit high infection risk, also should be considered as risky contacts and the corresponding risk of each area should be also re-assessed according to the number and the type of the individuals inside this area. Hence for an individual that is user of the application, defined as ``referencing point", and a given radius $r$ around him, in our approach we define the risk assessment of the area inside the radius $r$ denoted as $A(R)$ computed  as follows:

\begin{equation}
 \resizebox{0.90\hsize}{!}{$A(r)=\frac{\sum\limits_{i=1}^{n_{A}}w_{A}\cdot d_{A_i}+\sum\limits_{i=1}^{n_{B}}w_{B}\cdot d_{B_i}+\sum\limits_{i=1}^{n_{C}}w_{C}\cdot d_{C_i}+\sum\limits_{i=1}^{n_{D}}w_{D}\cdot d_{D_i}}{\sum\limits_{p=1}^{N}w_{A}\cdot \overline{D}},$}
\end{equation}

\noindent where $n_A$, $n_B$, $n_C$, and $n_D$ correspond to the cardinalities of the sets of the four categories of individuals, i.e., infected individuals, contacts of infected individuals, contacts of contacts, and random individuals, respectively,  $w_A$, $w_B$, $w_C$, and $w_D$ correspond to the weights of each category, i.e., infected, contacts of contacts and random individuals, $d_{X_i}$ corresponds to the distance between the referencing point and the $i-$th individual of category $X$, let $X_i$, $N$ corresponds to the total number of individuals (i.e., number of devices scanned nearby), and $\overline{D_A}$ corresponds to the average distance between the nearby detected individuals and the reference point, if the the nearby detected individuals where all infected. 

\noindent Respectively, for and arbitrary set of defined categories this equation in defined as:

\begin{equation}
A(r)=\frac{\sum\limits_{k=1}^{K}\sum\limits_{i=1}^{|C_k|}w_{C_k}\cdot d_{C_{k_i}}}{\sum\limits_{p=1}^{N}w_{C_1}\cdot \overline{D}},
\end{equation}

\noindent where, $K$ denotes the number of defined categories, and $|C_k|$ denotes the cardinality of the $k-$th category.

In our proposed approach, we distinguish five risk Classes, namely $A, B, C, D,$ and $E$, characterizing in ascending order the risk of an area regarding both the congestion in terms of the status of nearby Bluetooth enabled devices as also their congestion, based on the computation of $A(r)$ (see, Equation~1) as depicted in Table~\ref{TAB_CLS}.

{\renewcommand{\arraystretch}{1.6}
\begin{table}[h!]
\centering
\begin{tabular}{clc}
\textbf{CLASS} & \textbf{Risk Assessment} & \textbf{Assessment}  \\ \hline
\textbf{\color{ForestGreen}{A}} & \phantom{X}$0 \leq A(r) \leq 0.2$ & Very Low Risk \\ 
\textbf{\color{Blue}{B}}& $0.2 < A(r) \leq 0.4$ & Low Risk \\ 
\textbf{\color{Orange}{C}}& $0.4 < A(r) \leq 0.6$ & Medium Risk \\ 
\textbf{\color{Red}{D}} & $0.6 < A(r) \leq 0.8$ & High Risk \\ 
\textbf{\color{Purple}{E}}& $0.8 < A(r) \leq 1$\phantom{X} & Very High Risk \\ 
\hline
\end{tabular}
\vspace{0.05 in}
\caption{Area Risk Assessment Classes Regarding the Number of Nearby Devices belonging to Infected Individuals.}
\label{TAB_CLS}
\end{table}
}

\begin{figure}[t!]
\centering
		\includegraphics[scale=0.62]{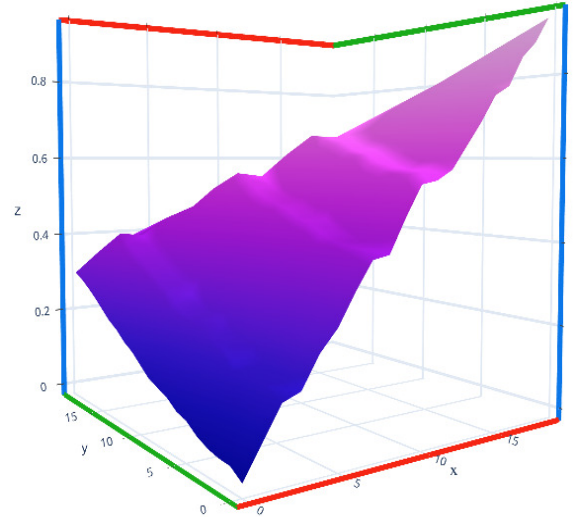}
	\caption{Evolution of the surface produced by the $A(r)$ curve for the categories of infected individuals and individuals that have been contacts of infected ones.}
	\label{fig:fig3}
\end{figure}

 In Figures~\ref{fig:fig2} and ~\ref{fig:fig3}, we demonstrate the achieved experimental results exhibited for the evaluation of the risk assessment $A(r)$ of an are of $100~m^2$ that includes $20$ individuals based on the cascade of the demographic data regarding the population density in Greece, and taking into account a radius of $10~m$ which corresponds approximately to the Bluetooth range. To this point, it is worth noting that throughout the experimental setup followed for the study of Equation 1, we performed a series of sequential experiments, since the total number of individuals surrounding the reference point individuals where randomly placed in different distances, as also belonging to different categories w.r.t to the combination that depicted each distribution of the set of individuals over the four categories. In the averaged results of Figure~\ref{fig:fig2} we utilized 4 categories, namely $C_A$, $C_B$, $C_C$, and $C_D$ corresponding to the sets of infected individuals, contacts of infected individuals, contacts of contacts, and random individuals, tuning respectively the corresponding weights as $w_A=0.7$, $w_B=0.2$, $w_C=0.09$, and $w_D=0.01$, respectively. To this point, we should note that the generic formula to compute all the possible combinations for a population sample of $N$ individuals over an arbitrary number of categories, let $P$, is computed as follows:


\begin{equation}\resizebox{0.9\hsize}{!}{$
\sum\limits_{K=1}^S\prod\limits_{M=1}^{P}{ {N-\sum\limits_{T=0}^M{C_T}} \choose {C_M }}:C_0=0,S \leq N,$ and $ S,P,N \in \mathbb{Z},$}
\end{equation}

\smallskip
\noindent where, $S$ is the total number of detected individuals, i.e., the sum of the cardinals of all the sets of categories, $P$ is the number of categories, $N$ is the total number of nearby individuals, $C_M$ and $C_T$ is the number of selected individuals from category $M$ and $T$, respectively. 

In the corresponding plot of Figure~\ref{fig:fig2}, the $x-$axis represents the index of each combination regarding the arrangement of the total of $20$ individuals  i.e., $|C_A|+|C_B|+|C_C|+|C_D|=20$, in the corresponding four categories, i.e., $C_A, C_B, C_C,$ and $C_D$, including cases such as $|C_A|=x,|C_B|=y,|C_C|=z,$ and $|C_D|=w$, where it holds that $0\leq x+y+z+w \leq 20$, while the $y-axis$ corresponds to the value of $A(r)$ for each one of the corresponding combinations in the distribution of the total $20$ individuals over the four categories. Hence, the index value  $1$ corresponds to the case where $|C_A|=20,$ and $|C_B|=|C_C|=|C_D|=0$, while the index value $10627$ corresponds to the case where $|C_A|=|C_B|=|C_C|=|C_D|=0$, and the index values in that range correspond to all the other combinations that depict the distribution on the cardinalities of the sets $C_A, C_B, C_D,$ and $C_D$, such that $|C_A|=x,|C_B|=y,|C_C|=z,|C_D|=w: 0\leq x+y+z+w \leq 20$.

In Figure~\ref{fig:fig3} we represent the evolution of the surface produced by the curve regarding the most important categories of individuals, namely $C_A$ and $C_B$, i.e., the individuals registered as infected $x-axis$ and individuals that have been contacts of individuals previously registered as infected $y-axis$, respectively, where the $z-axis$ depicts the value of $A(r)$ for each combination of distribution of individuals over $C_A$ and $C_B$, w.r.t. the settings followed in the previous experiments.

\subsection{System Deployment}
The application is intended to be deployed in any Bluetooth enabled smart-device that has the capability of running an app. The leading smart-phone device operating systems are Android (72.19\% as of April 2021) and iOS (27\% as of April 2021)~\cite{statcounter}. The application is intended to be distributed via the device's according application store (Play Store / App store). The reason behind the choice of the application stores is mainly for convenience as most of smart devices users are accustomed to the stores as the main distributors of software for their devices. The proposed application will receive regular updates regarding its UI/UX and any feature necessary to tackle more efficiently the proactive prevention of the spread deploying an efficient contact tracing procedure. Our System has two actors. The first is the user of the application and the second is the medical staff that uses the web application. Next, we enumerate the basic procedures utilized for the deployment of the SARiSa application for the proactive control of the spread of airborne virus SARS-CoV-2. The procedures are discussed over their deployment concerning the immediate users i.e., the individuals, as also the authorized medical staff that operates on behalf of the application.  

\bigskip
\noindent\textbf{1) User Registration.}
The registration procedure allows the application user to set up an account and successfully register his/her information in the database allowing the successfully registered user to deploy the application (see, Procedure~1). In order for the unregistered user to start the registration procedure it is required by the back-end system that the user holds an One-Time-Code (OTC) provided by the authorized medical staff. On registration, the application user fills the required information alongside the corresponding credentials in a registration form appeared inside the GUI of the application, and the server-side application attest the validity of the inserted credentials before allowing the user to deploy the application through his Bluetooth enabled smart mobile device.

{\renewcommand{\arraystretch}{1.6} 
\begin{table}[h!]
\begin{tabular}{p{8.2cm}l} 
\hline
 \textbf{Description:} This procedure allows the SARiSa application user to set up an account and register the required information in the database.\\
 \hline
  \textbf{Prerequisites:} The user has a unique OTC\\
  \textbf{Actors:} Application User\\
  \textbf{Event Flow:}\\
 \textbf{1.} The \textit{Application User} presses the REGISTER button\\
 \textbf{2.} A registration form appears\\
 \textbf{3.} The \textit{Application User} enters his/her information\\
 \textbf{4.} The \textit{Application User} presses the APPLY button\\
 \hspace{3mm}\textbf{4.1.a} Check information validity (Proceed to Step 5)\\   
 \hspace{3mm}\textbf{4.1.b} The \textit{Application User}'s information is  not valid\\
 \hspace{10mm}Re-enter user information (Return to Step 3)\\
 \textbf{5.} The \textit{Application User} logs in the application starting its deployment.\\
 \hline
\end{tabular}
\vspace{0.05 in}
\caption*{\small{\textbf{Procedure 1}: User Registration}}
\label{Tab1}
\end{table}
}

\noindent\textbf{2) Registering Verified Users.}
The procedure of attesting the users that are registered in the application by assuring that their status (i.e., if they have been detected as infected by authorized medical staff) is also certified, has its basis on the utilization of an One-Time-Code number that is generated by the account creation in the data-base and is given to the authorized medical staff to be shown to the individual who need to complete his registration.  Once the authorized medical staff has initiated the account creation for an individual the system returns an OTC (see, Figure~\ref{fig:fig1}) to the authorized medical staff which is then passed to the individual that wants to be registered (see, Procedure~2). The back-end procedures are responsible to attest that the user registration procedure contains this unique OTC, and hence the registration of the user as also the status of his health, i.e., susceptible/infected, or recovered are also verified since the procedure has been initiated by the authorized medical staff. 

{\renewcommand{\arraystretch}{1.6} 
\begin{table}[h!]
\begin{tabular}{p{8.2cm}l} 
\hline
 \textbf{Description:} This procedure allows an individual to be registered in the application utilizing a valid OTC obtained by the authorized medical staff proceeding with the deployment of the SARiSa application.\\
 \hline
 \textbf{Prerequisites:} Not Reported\\
 \textbf{Actors:} Application User\\
  \textbf{Event Flow:}\\
 \textbf{1.} The \textit{Application User} presses the ENTER OTC button\\
 \textbf{2.} A text area appears\\
 \textbf{3.} The \textit{Application User} enters the OTC\\
 \textbf{4.} The \textit{Application User} select the status\\
 \textbf{5.} The \textit{Application User} presses the APPLY button\\
 \hspace{3mm}\textbf{5.1.a} Check \textit{Application User}'s OTC validity (Proceed to Step 6)\\
 \hspace{3mm}\textbf{5.1.b} The \textit{Application User}'s OTC is checked for validity and fails\\ 
 \hspace{10mm}Re-enter the OTC (Return to Step 3)\\
 \textbf{6.} The \textit{Application User}'s status changes\\
 \hline
 \end{tabular}
\vspace{0.05 in}
\caption*{\small{\textbf{Procedure 2}: User Verification using the One-Time-Code (OTC)}}
\label{Tab2}
\end{table}
}

\bigskip
\noindent\textbf{3) Scanning for Nearby Infected Individuals.}
The core component of our proposed application for the proactive control of airborne diseases, SARiSsa, has its basis on the detection and evaluation of the risk exhibited by the nearby contacts located in the area around a user of the application. The main target of SARiSsa application is to proactively prevent the infection of a susceptible individual by the nearby potentially infected individuals, utilizing the Bluetooth to detect the nearby Bluetooth enabled devices and retrieving their status registered inside their profile in the application respectively.  Hence, the procedure of nearby device scanning, detects whether there exist nearby Bluetooth enabled devices that have installed the application and by retrieving the relative information regarding on whether their status has been set to infected, or whether they are immediate contacts of infected individuals, proceeding to the risk assessment of the area characterizing this area by the corresponding risk class, achieving thus to inform the user and proactively protect him from a probable infection (see, Procedure~3). In contrast to the applications proposed so far, SARiSsa, not only informs its user whether they have been a contact of an individual later detected as infected but more importantly proactively protects its users by informing them in real time regarding the risk of a congested area before they stay for more time close to potentially risky contacts, i.e., infected individuals or contacts of infected individuals, or even individuals that have remained for longer time in areas with high infection risk. 

{\renewcommand{\arraystretch}{1.6} 
\begin{table}[h!]
\begin{tabular}{p{8.2cm}l} 
\hline
 \textbf{Description:} This procedure allows the application user to scan for nearby  Bluetooth enabled devices and assess the risk of the user catching the virus.\\
 \hline
 \textbf{Prerequisites:} Not Reported\\
  \textbf{Actors:} Application User\\
  \textbf{Event Flow:}\\
 \textbf{1.} The \textit{Application User} starts the SCAN procedure\\
 \textbf{2.} The system scans the environment for neighboring devices\\
 \textbf{3.} The system assesses the risk\\
 \hspace{3mm} \textbf{3.1} Detection of nearby Bluetooth Enabled Devices\\
 \hspace{3mm} \textbf{3.2} Search in Data-Base for registered UUIDs \\
 \hspace{3mm} \textbf{3.3} Retrieve individual status for these UUIDs\\
 \hspace{3mm} \textbf{3.4} Conduct Area Risk Assessment $A(r)$.\\
 \hspace{3mm} \textbf{3.5} Perform risk evaluation.\\
\textbf{4.} Inform user about the Risk Class of surrounding area\\
 \hline
\end{tabular}
\vspace{0.05 in}
\caption*{\small{\textbf{Procedure 3}: Nearby Device Scanning and Notification}}
\label{Tab3}
\end{table}
}

Regarding the confidentiality and the privacy properties required for the sensitivity of the personal health data, the deployment of the SARiSsa application is based on the exchanged values of {\tt UUID} and other Bluetooth related information that are shared between the nearby devices as also being stored into the systems data-base, that are all recorder by their hash values being also uniquely traceable, while on the other hand cannot indicate in any way the identity of an individual.

\bigskip
\noindent\textbf{4) Web-based User Registration.}
Another important procedure provided by the proposed application is the one of user registration deployed by the authorized medical staff (see, Procedure~4). Throughout this procedure, once an individual has been detected as infected throughout the conduction of a medical testing, i.e., rapid tests or a Polymerase Chain Reaction (PCR) tests, the authorized medical staff inserts a specific record in the system, generating a Ont-Time-Code, that will be given to the infected individual in order to use it on his status update though his application interface in order to securely ensure that his registration as infected has been verified by an authorized medical staff. Next, once the OTC has been generated by the system, the authorized medical staff gives this code to the user of the application in order to utilize it as for his authorization in the procedure of status update to be validated. Next, once the OTC has been checked by the back-end of the application and ensuring the validity of the inserted credentials, or been asked to re-enter a valid OTC generated by an authorized medical staff otherwise, the user can select to update his status to ``infected", proceeding finally to the completion of the procedure by get the necessary information by the system that his status update has been made successfully.

{\renewcommand{\arraystretch}{1.6} 
\begin{table}[h!]
\begin{tabular}{p{8.2cm}l} 
\hline
 \textbf{Description:} This procedure allows the medical staff to register a user's information into the database in order for the user to update a user's status . \\
 \hline
 \textbf{Prerequisites:} Not Reported\\
  \textbf{Actors:} Medical Staff, Application User\\
 \textbf{Event Flow:}\\
 \textbf{1.} The \textit{Medical Staff} presses the ``register" button\\
 \textbf{2.} The system generates a unique OTC\\
 \textbf{3.} A message appears to the \textit{Medical Staff} displaying the OTC\\
 \textbf{4.} \textit{Application User} enables the ``Update Form"\\
 \textbf{5.} \textit{Application User} selects ``User Status"\\
 \textbf{6.} \textit{Application User} inserts the OTC \\
 \textbf{7.} \textit{Application User} applies changes\\
 \hspace{3mm}\textbf{7.1.a} Check OTC validity (Proceed to Step 8)\\
 \hspace{3mm}\textbf{7.1.b} The OTC is checked for validity and fails\\
 \hspace{10mm}Re-enter the OTC (Return to Step 6)\\
 \textbf{8.} Finalizing ``Status Update" updating the DB\\
 \textbf{9.} Inform user for the successful update\\
 \hline
\end{tabular}
\vspace{0.05 in}
\caption*{\small{\textbf{Procedure 4}: Web-based User Status Update}}
\label{Tab5}
\end{table}
}

\section{Spread Control.}
In this section we present the deployment of our proposed model concerning the utilization of the application SARiSsa, providing some experimental results achieved through a series of simulation experiments that prove the potentials of the proposed algorithmic background regarding the deployment of the contact tracing procedure and its effectiveness in pandemic mitigation.

\begin{figure*}[t!]
 \begin{minipage}{\linewidth}
    \begin{minipage}{0.45\linewidth}
        \includegraphics[scale=0.38]{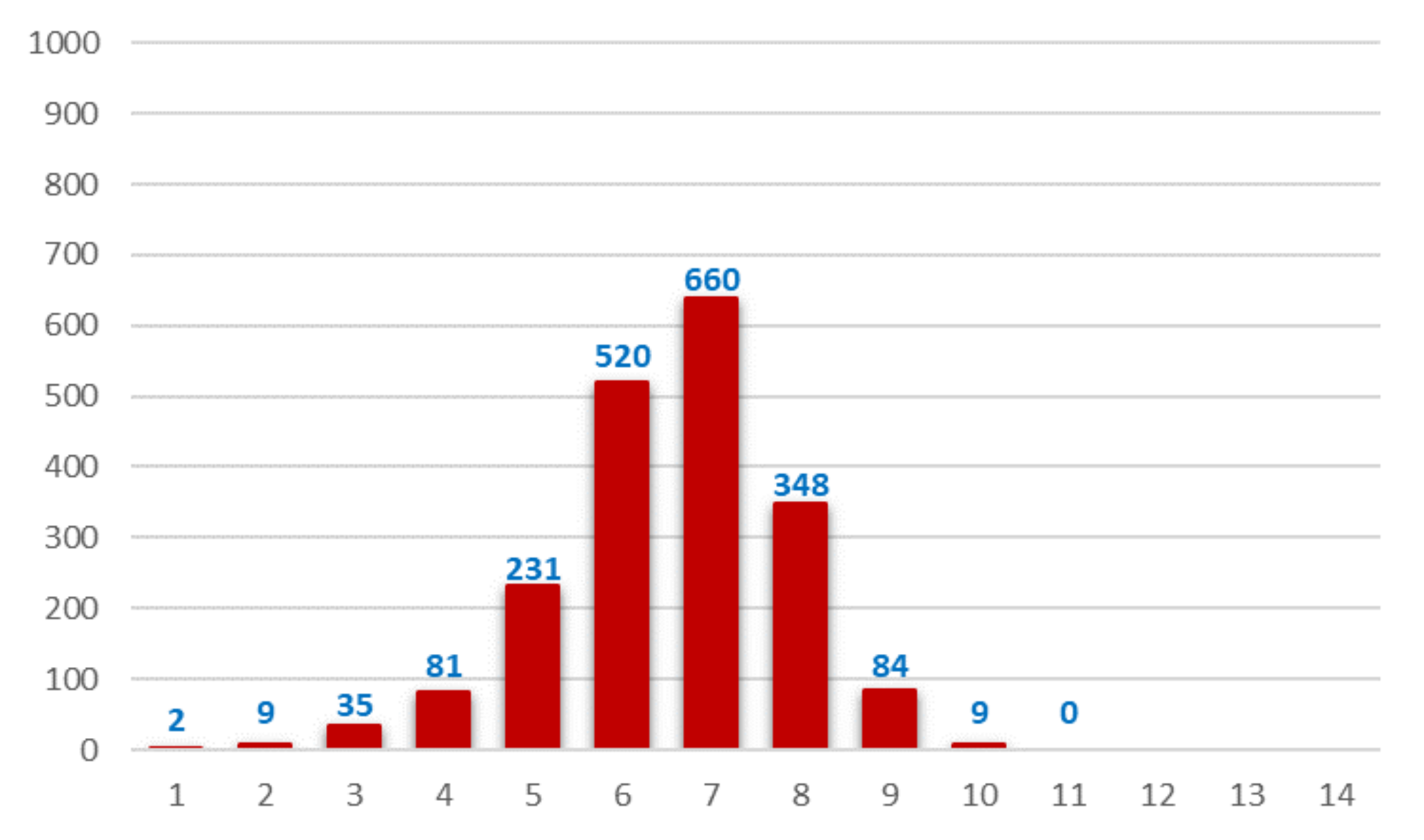} \\
        \vspace{0.05 in}\tiny{\phantom{xxxxxxxxxxxxxxxxxxxxxxxx}(a) Daily infection cases without the deployment of SARiSsa.}
    \end{minipage} \hspace{0.4 in}
    \begin{minipage}{0.45\linewidth}
       \includegraphics[scale=0.4]{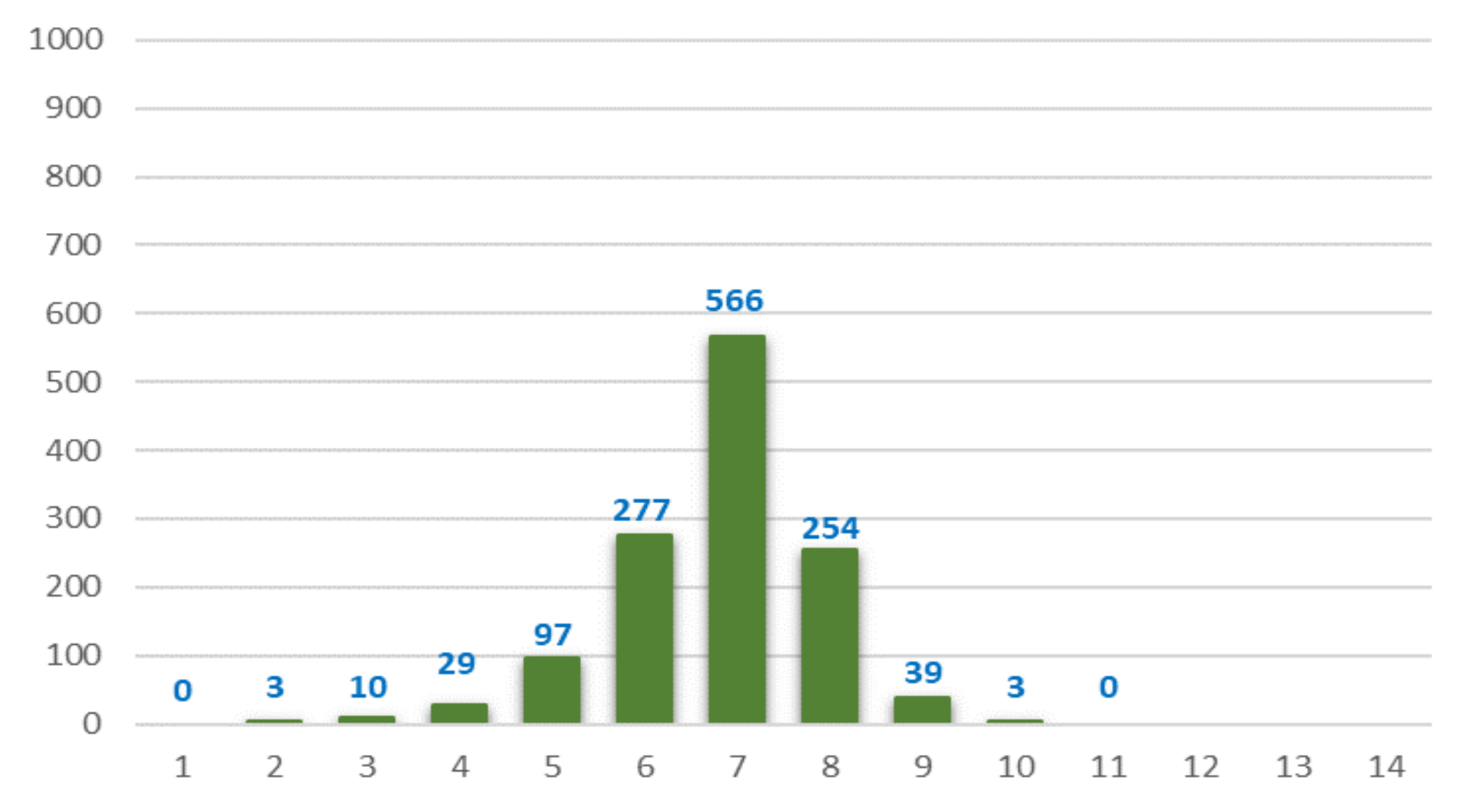}\\
        \vspace{0.05 in}\tiny{\phantom{xxxxxxxxxxxxxxxxxxxxxxxx}(b) Daily infection cases with the deployment of the SARiSsa.}
    \end{minipage}

     \smallskip\smallskip  
    \caption{Averaged daily infection cases regarding the spread of an airborne pathogen taking to no countermeasure to control the spread (a), and by deploying the SARiSsa smart application for contact tracing.}
\label{App-Results}
\end{minipage}
\end{figure*}	

	\subsection{Agile Deployment of SARiSsa Contact Tracing Mobile Application}
	Assuming that the user has already downloaded the app to his smart phone,
and successfully register meaning that the database has a copy of his/hers information.
In the first step the user opens the app and following that a back-end service is scanning
nearby Bluetooth devices. The aim of the scan is to let the app have access to individual
Bluetooth {\tt UUID}'s as they represent a unique identifier for each device.  The next
step is to fetch the {\tt UUID}'s to the user database and lets say that those {\tt UUID}'s are the daily contacts of an individual. The database then retrieves the contacts
related to the user of the app, and those data are given to the status fetcher
that process them and then hand them over to the risk assessment module to 
compute the infection risk. The risk assessment from the other hand must notify
the app through the notifier module about the infection status. Finally the app
informs the user about his infection status.

Next, focusing on the event notifier module and what triggers it. Except from the 
risk assessment module, events can be triggered by status checker and the combination
of contact notifier and create risk notification modules. Status checker is a module
that runs in the background and  periodically gets the user status
from the database and check if this status have been altered, for enable a negative COVID-19
user became positive.if the status becomes positive the module immediately calls event notifier to inform the user about the status. If the status checker returns a negative result means that a risk assessment has to be made. So the status checker calls the contact list creator so as to create a contact list of the current date contacts as for contacts have been made at least 3 days before. Then based on the status of those contacts a risk estimation
is produced. Along with the risk assessment also the suspicious contacts are hand over. Finally the user and users involved are notified about the infection risk.

Moreover, another possible scenario is that we have a patient that has not been registered yet to the app. The obvious problem with this scenario is that we do not have any information about the user. So in our proposed architecture a web app is used by the health center
personnel to update the user status. Firstly lets assume that the medical staff is logged-in
the web app and requests a code unique to the patient. The code is generated and hand
over to the patient as well is stored to the user database. The patient puts
the code to the web app and automatically his status is updated. Along this process
we assume that the web app retrieves patients Bluetooth {\tt UUID} and with a method
identical to the status checker module after a negative status, the web app produces
a risk assessment so as to inform users that have patients {\tt UUID} in their contacts
that they are at risk because of the positive COVID-19 outcome test of the patient.

	\subsection{Effectiveness of Contact Tracing}
In order to evaluate the effectiveness of the proposed contact tracing application, we performed a series of simulation experiments utilizing the integrated framework proposed in \cite{NiPo}. We transformed some characteristics of the simulation framework for the investigation of the spread of malicious software between mobile devices, meeting the demands of airborne transmitted diseases.

 Throughout the series of simulation experiments, we consider the utilization of the SARiSsa mobile application to mitigate the spread of the airborne virus (in our case the SARS-CoV-2) and compare the results against a potential spread that could occur without the utilization of this application. To this point, we ought to notice that throughout our simulation experiments we consider the basic deployment of a tracing application deployed through Bluetooth enabled devices, focus to investigate the positive effect that could exhibit in the contact tracing procedure.

Regarding the manipulation of the infected and infectious population, it is worth noting that through our simulation experiments we consider any infected individual to be set in quarantine for ten days, starting on two days after the symptoms onset. Due to the contact monitoring capabilities of the proposed smart application, when an infectious individual is set on quarantine, his most frequent contacts from the last two days will be notified via the application and will be also set on quarantine for ten days as well. 

For the experimental results achieved from a series of simulation experiments depicted in Figure~\ref{App-Results}, we consider a sample of 2000 individuals consisting of $1$ infected individual and $1999$ susceptible ones, distinguishing in each experiments series two main cases, with respect to the lack in the deployment of a smart application for the contact tracing procedure, as also in presence of such an application, respectively, see Figure~\ref{App-Results}~(a) and Figure~\ref{App-Results}~(b). To this point we should notice that throughout our simulation experiments we assumed that no safety distance is retained between the individuals, nor the utilization of FFP1/FFP2 face-mask protection. As presented in Figure~\ref{App-Results}~(b) when compared to Figure~\ref{App-Results}~(a), there is a significant difference in the infected population regarding the case where no counter-measure is taken, i.e., no smart application deployed for the contact tracing procedure. Additionally, in contrast to the experiment after the deployment of the SARiSsa contact tracing application, in the corresponding experiment in which no smart application is deployed for the contact tracing procedure all the individuals are infected leading to a pandemic at day $10$. Moreover, regarding the experiment where the deployment of the SARiSsa contact tracing application is present, the daily infection cases show a linear decreasing rate until the number of new infection cases is almost eliminated. Finally, as we can observe in Figure~\ref{App-Results}~(b)), the pandemic has been effectively mitigated thanks to the use of the SARiSsa mobile application, where a total of $1278$ individuals got infected instead of the $1979$ that would got infected if no countermeasure was deployed.

\section{Conclusion}

In this section we conclude our work, discuss the potentials and the improvements that we plan for the proposes SARiSsa contact tracing mobile application  alongside our aims for future research.

	\subsection{Discussion}
One of the main concerns over the procedure of contact tracing is primarily to locate as soon as possible the immediate contacts recently made by an individual that has been declared as infected (i.e., the contacts made the past two dates before the appearance of the first symptoms, ore the testing date). To this direction, we have designed an efficient contact tracing algorithm (see, Algorithm 1) intended to extend this procedure to a wider range of individuals inducing the current contacts of the contacts an infected individual made two days ago, providing hence a proactive safety against the spread of the diseases by nearby individuals. The intuition behind the design of this algorithm is that if the infected individual has already infected some of the contacts he made two days ago, then these contacts currently should be infectious, and hence the contacts they made the current date (i.e., second degree contacts) should also be set to quarantine as potentially infected. Hence, for each individual we deploy the proposed algorithm for contact tracing  utilizing its contact list, $Two\_D\_CL$ that includes its contacts from the lat two days. We proceed by our intuition tht the contacts he made exactly two days before the symptoms appeared, are more like to have already an adequate viral load in the current date being infectious enough to transmit the disease to the contacts the made in the current date. So for each individual we select from his contact list the contacts he had two days before the date that the symptoms appeared, and for each one of these contacts we scan its contact list $tmp\_Co\_CL$ and select the contact they made the current date (i.e., contacts of contacts, or for short, Co-Contacts. We store all these individuals in a list, $Co\_CL$, and return it as output of the algorithm in order for our system to notify them through the mobile application as to be isolated.

\begin{algorithm}[h!]
 \caption{Contact Tracing Algorithm.}
  \small{
 \KwData{List of Contacts of the Last $2$ Days: $Two\_D\_CL$}
 \KwResult{List of Co-Contacts: $Co\_CL$}
 List $Co\_CL \leftarrow null$\;
 List $tmp\_Co\_CL \leftarrow null$\;
 var $index \leftarrow 0$\;
 \While{$Two\_D\_CL(index)\neq null$}{
  \If{$Two\_D\_CL(index).date\_of\_contact-Current\_Date()=2$}{
       $tmp\_Co\_CL \leftarrow Two\_D\_CL(index).Two\_D\_CL$\;
       $tmp\_index \leftarrow 0$\;
       \While{$tmp\_Co\_CL(tmp\_index)\neq null$}{
       	     \If{$tmp\_Co\_CL(index).date\_of\_contact-Current\_Date()=0$}{
       	          $Co\_CL.add(tmp\_Co\_CL(tmp\_index).id)$\;
       	     }
       	     $tmp\_index++$\;
       	     \If{$tmp\_Co\_CL(index)=null$}{
                  $break$\;
             }
       }
       
  $Co\_CL.add(Two\_D\_CL(index).id)$\;
  }
  $index++$\;   
  \If{$Two\_D\_CL(index)= null$}{
       $break$\;
  }
 }}
\end{algorithm}

	\subsection{Remarks and Future Research}
In this work we propose an open architecture for the development of contact tracing applications deployed on smart mobile devices to perform efficiently the contact tracing procedure. We propose the development of such a mobile application that is based on the interaction among nearby Bluetooth enabled devices, the so-called SARiSsa, that incorporates a set of efficient and effective procedures to scan, detect and informs the user about the risk of an area regarding the infectious population or, indirectly if they are been contacts of infected ones, providing a proactive protection against the spread of SARS-CoV-2 or any other airborne disease.  Finally, beyond the open architecture proposed in this work, we also propose an efficient and effective algorithm to augment ad extent the contact tracing procedure,designed to actively prevent the spread from potentially infected individuals that the traditional contact trace procedures so far omitted.

	Concerning our aims for our future research we plan to investigate the design of more effective contact tracing procedures. Additionally, we focus on the development of more robust and widely acceptable security methods and lightweight cryptographic techniques that will allow the wider acceptance of the application dealing with the issues of data privacy and confidentiality of health care information against the functionality and the efficiency required for mobile tracing applications. Finally, we plan to integrate the development of the SARiSsa mobile application for contact tracing and perform a real-life scenario evaluation as soon as the data privacy issues are fulfilled.


\end{document}